\documentclass[a4paper,twoside]{article}

\usepackage{epsfig}
\usepackage{subcaption}
\usepackage{calc}
\usepackage{amssymb}
\usepackage{amstext}
\usepackage{amsmath}
\usepackage{amsthm}
\usepackage{multicol}
\usepackage{pslatex}
\usepackage{apalike}
\usepackage{algorithm2e}
\usepackage{minted}
\usepackage[bottom]{footmisc}
\usepackage{SCITEPRESS}     % Please add other packages that you may need BEFORE the SCITEPRESS.sty package.
\begin{document}

\title{SCART: Simulation of Cyber Attacks for Real-Time}

\author{\authorname{Eliron Rahimi\sup{1}\orcidAuthor{0009-0007-4424-2123}, Kfir Girstein\sup{2}\orcidAuthor{0009-0003-4976-0561} , Roman Malits\sup{3}\orcidAuthor{0000-0001-8989-8125} and Avi Mendelson\sup{3}\orcidAuthor{0000-0003-4274-6866}}
\affiliation{\sup{1}Computer Science Department, University of Haifa, Haifa, Israel}
\affiliation{\sup{2}Department of Electronic Engineering, Technion Institute of Technology, Haifa, Israel}
\affiliation{\sup{3}Department of Computer Science, Technion Institute of Technology, Haifa, Israel}
}


\abstract{Real-time systems are essential for promptly responding to external stimuli and completing tasks within predefined time constraints. Ensuring high reliability and robust security in these systems is therefore critical. This requires addressing reliability-related events, such as sensor failures and subsystem malfunctions, as well as cybersecurity threats.
This paper introduces a novel cyber-attack simulation infrastructure designed to enhance simulation environments for real-time systems. The proposed infrastructure integrates reliability-oriented events and sophisticated cybersecurity attacks, including those targeting single or multiple sensors.
We present the SCART framework and dataset, addressing a central challenge in real-time systems: the lack of scalable testing environments to assess the impact of cyber-attacks on critical systems and evaluate the effectiveness of defensive mechanisms. This limitation arises from the inherent risks of executing attacks or inducing malfunctions in operational systems. By leveraging simulation-based capabilities, the framework generates training and testing data for data-driven approaches, such as machine learning, which are otherwise difficult to train or validate under live conditions.
This development enables the exploration of innovative methodologies to strengthen the resilience of real-time systems against cyber-attacks. The comprehensive functionalities of the proposed infrastructure improve the accuracy and security of critical systems while fostering the creation of advanced algorithms. These advancements hold the potential to significantly enhance anomaly detection in real-time systems and fortify their defenses against cyber threats. Our code is available at https://github.com/kfirgirstein/SCART.}

\onecolumn
\maketitle
\normalsize
\setcounter{footnote}{0}
\vfill
\footnotetext{This research was supported in part by the zero-trust project, TII -- Technology Innovation Institute, Abu-Dhabi.}

\section{\uppercase{Introduction}}
\label{sec:introduction}

 Real-time systems typically function as reactive systems, promptly responding to stimuli and completing tasks within predefined time limits. The correctness of these systems depends on both the accuracy of their outputs and their adherence to specified timing constraints \cite{laplante2004real}. As the number of transistors on a single chip continues to increase exponentially, the complexity of such systems grows accordingly \cite{Schoeberl2019ArchitectureOC}. Today, most systems are composed of multiple components, each of which may be designed and tested by different teams or even by separate companies \cite{inproceedings}. As a result, modern systems have become sufficiently complex that it is nearly infeasible to exhaustively test and verify all possible usage models and execution paths. This introduces vulnerabilities to errors, incomplete specifications, and cyber-attacks. 

Real-time embedded systems are susceptible to a variety of attacks \cite{papp2015embedded}, rendering their security a significant concern. While various strategies have been proposed to mitigate these issues over time, there remains an urgent need for enhanced tools to support the development of new protective measures and to test a system’s robustness against different attack vectors. 

Simulators, particularly digital twin systems \cite{he2021digital}, provide an effective means for users to validate the behavior of proposed schemes and test new algorithms under diverse conditions and situations \cite{sargent2010verification}. While some simulators allow for the analysis of the effects of faults, such as sensor damage or component malfunctions, they typically focus on systems operating under normal conditions rather than simulating failure scenarios or security attacks. These attacks may range from simple, single-point failures to more intricate, multi-vector threats.

This paper introduces the SCART layer, a novel abstract solution designed to augment existing real-time system simulators. SCART augments these simulators by enabling the detection of faults and cyber-attacks, allowing both types of threats to be incorporated into simulated environments. SCART's architecture seamlessly integrates a wide range of cyber threats into simulations, including simple attacks like single-sensor failures and more complex security breaches that affect multiple sensors at the System-on-Chip (SoC) level or across the entire controller. 

To assess SCART's efficacy, a series of experiments were conducted with various attack scenarios on multiple real-time system simulators, including Gazebo, Airsim, and jMAVSim.  These experiments involved simulating specific flight paths and evaluating 2,048 different attack configurations involving sensor manipulations and combinations of attacks.  We evaluated the performance of existing anomaly detection algorithms using established benchmark methods \cite{blazquez2021review}. Notably, these algorithms identified approximately 86\% of non-anomalous instances and 78\% of anomalous instances, underscoring the importance and effectiveness of this research. The experiments provided valuable insights into the reliability and utility of SCART as a tool for generating realistic cyber-attack scenarios.

\subsection{Contributions}
The key contributions of this paper are as follows:
\begin{itemize}
    \item Introducing the SCART environment and demonstrating its efficiency and capabilities.  
    \item Presenting an abstract framework architecture for integrating a cyber-attack layer into real-time system simulations.
    \item Providing a proof of concept for integrating the cyber-attack layer into a digital twin simulation for PX4 drones.
    \item Developing a multi-sensor time-series dataset for training algorithms to detect anomalies.
\end{itemize}

The remainder of this paper is structured as follows: Section \ref{Background} provides the necessary background for understanding the work, while Section \ref{related_works} reviews related research. In Section \ref{mainSection1}, we present the SCART architecture and a set of attack scenarios with different implementation approaches. Section \ref{mainSection2} describes the dataset generation process using our framework. We evaluate our work in Section \ref{mainSection3}, and finally, we present our conclusions in Section \ref{discussion}.

\section{\uppercase{Background}}
\label{Background}
\subsection{Simulations}

Simulation is a vital tool in the design and testing of control systems. It allows users to evaluate the performance of various schemes and algorithms under different conditions. Simulators are typically categorized into two main types: \textbf{functional simulators}, which prioritize efficiency but may sacrifice precise timing accuracy and \textbf{performance simulators}, which, although slower, offer cycle-accurate results.

Certain simulators also provide advanced sensor simulations and support both \textbf{Software-in-the-Loop (SITL)} and \textbf{Hardware-in-the-Loop (HITL)} simulations. In SITL simulations, the entire system is simulated using software on a computer, whereas HITL simulations involve connecting physical hardware to the computer to mimic the system’s behavior. Notable examples include \textbf{Gazebo} \cite{kaur2021survey}, \textbf{jMAVSim} \cite{drones6090261}, and \textbf{Airsim} \cite{airsim2017fsr}, which are widely used for testing robotic and autonomous flight algorithms. Additionally, Vehicle-focused simulators such as \textbf{CarSim} and \textbf{X-Plane} are also  commonly used \cite{kaur2021survey}.

Simulators based on the PX4 autopilot flight-control architecture \cite{7140074} are particularly effective for vehicle simulations. The MAVLink API simulator facilitates communication by transmitting sensor data from the simulation environment to the PX4 and relaying engine and actuator values from flight code to the simulated vehicle. This capability enables real-world-like interactions with the simulated vehicle, allowing users to control it via \textbf{QGroundControl}, an offboard API, or even a radio controller/game board. These simulators also provide capabilities for uploading waypoints and specifying additional task parameters using Python packages, enhancing their flexibility for research and development.

\begin{figure}[!t]
    \centering
    \begin{subfigure}[b]{0.32\linewidth}
        \includegraphics[width=\linewidth,height=0.1\textheight]{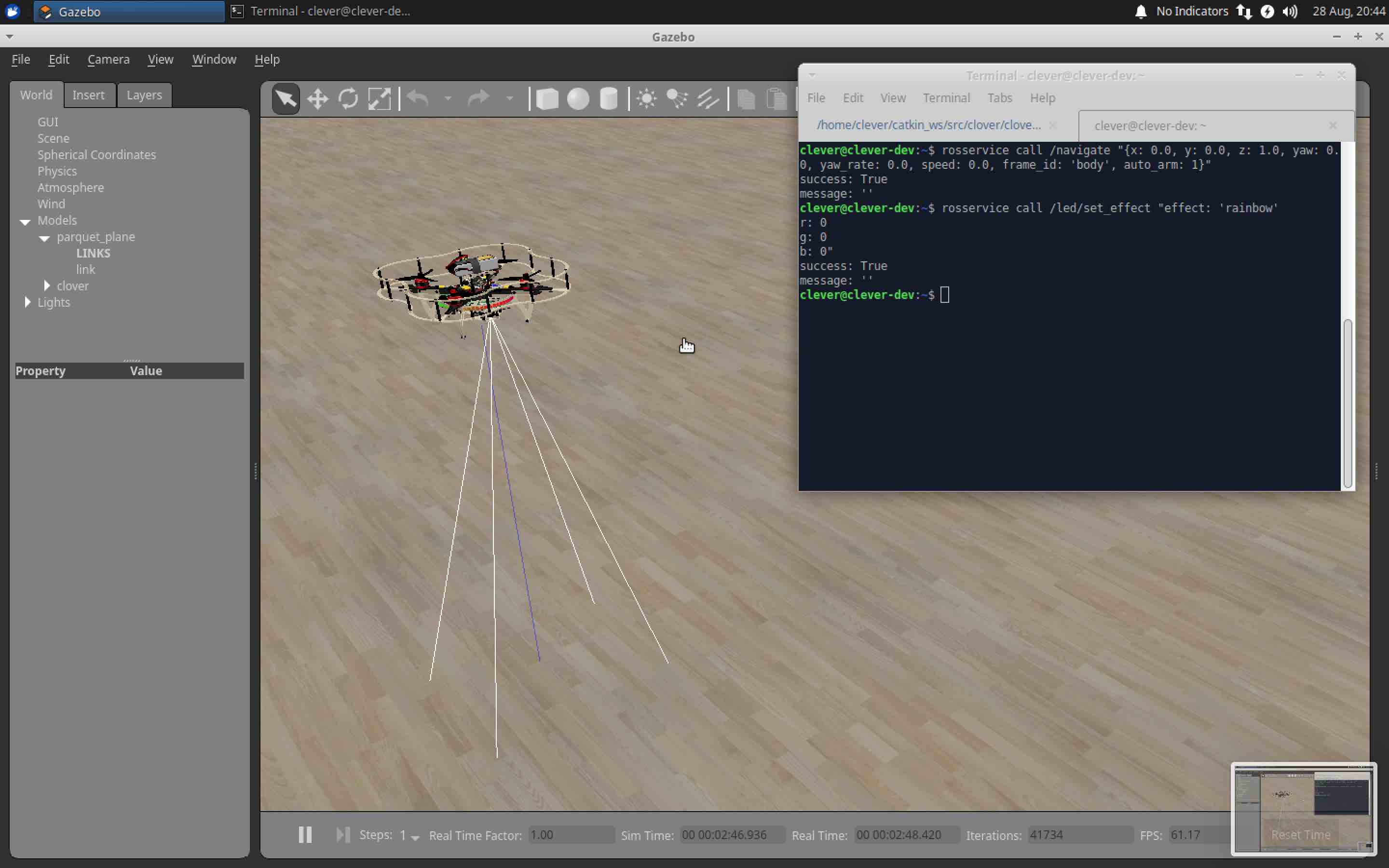}
        \caption{Gazebo.}
    \end{subfigure}
    \hfill
    \begin{subfigure}[b]{0.32\linewidth}
        \includegraphics[width=\linewidth,height=0.1\textheight]{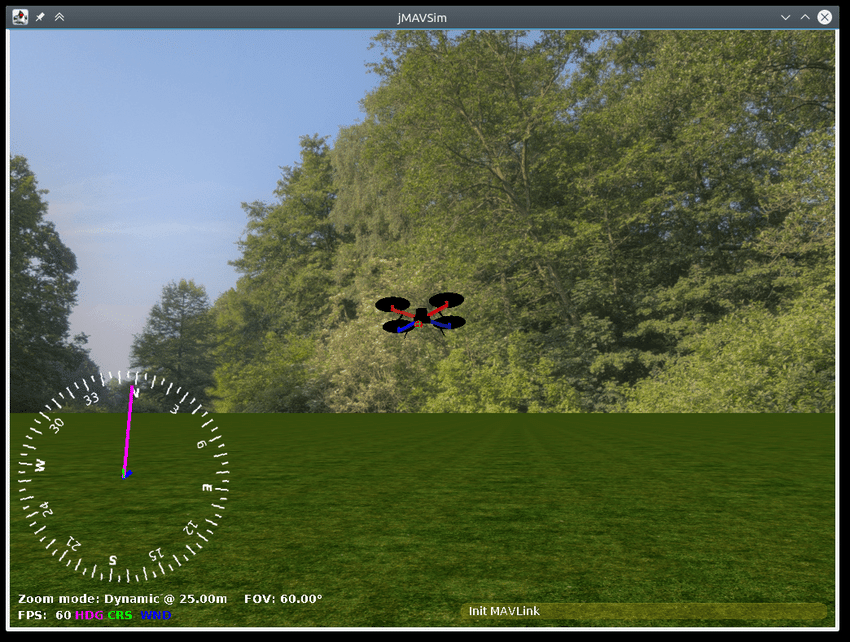}
        \caption{jMAVSim.}
    \end{subfigure}
    \hfill
    \begin{subfigure}[b]{0.32\linewidth}
        \includegraphics[width=\linewidth,height=0.1\textheight]{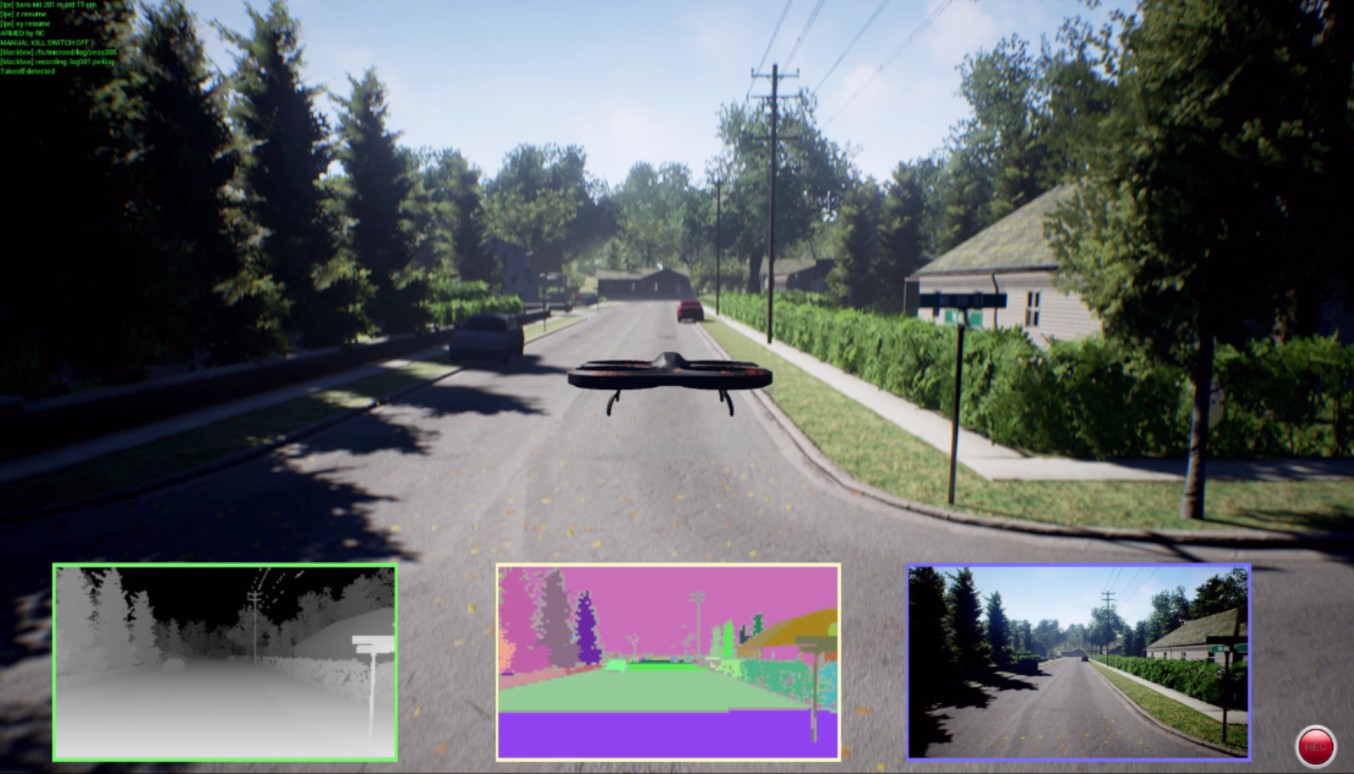}
        \caption{Airsim.}
    \end{subfigure}
    \caption{Snapshots from vehicle simulators: Gazebo, jMAVSim, and Airsim.}
    \label{fig:simulators}
\end{figure}

\subsection{Embedded System Threats}
\label{StealthThreats}

Embedded systems face numerous security threats that can compromise their integrity and potentially endanger their users. These threats include sensor-based attacks, which manipulate sensors to falsify data or block access to legitimate information, such as GPS spoofing and jamming \cite{humphreys2012statement}. Additionally, remote exploitation involves exploiting vulnerabilities to trigger unexpected behavior, as exemplified by buffer overflows \cite{lhee2002type}. In contrast, direct physical injection refers to tampering with hardware devices through physical access or injecting vulnerabilities manually \cite{parkinson2017cyber}. Supply chain attacks also pose a potential risk, where hardware, firmware, software, or system information may be maliciously compromised to steal, counterfeit, disrupt, or compromise device functionality. Notable examples of supply chain attacks include the Stuxnet attack \cite{7360168}, one of the most infamous cyber-warfare weapons, and the Hardware Trojan Horse (HTH) \cite{parkinson2017cyber}, a malevolent modification of the circuitry of an integrated circuit. These attacks have severe consequences, including the loss of sensitive information, unauthorized remote access, and even physical harm, highlighting the critical need to promptly identify and address these security threats to ensure the safety and security of embedded systems.

\section{\uppercase{Related work}}
\label{related_works}

Creating anomalies in time-series data can be achieved through either data generation or threat injection methods.

\textbf{Data generation techniques}, such as the TSAGen tool proposed by Wang et al. \cite{wang2021tsagen}, generate synthetic time-series data containing anomalies. While these methods are useful, synthetic data often fails to accurately represent anomalies caused by real-world phenomena, particularly those originating from cyberattacks.Consequently, anomaly detection algorithms trained solely on synthetic data may struggle to identify anomalies in real systems with high reliability.

\textbf{Threat injection methods}, on the other hand, introduce simulated cyber threats to produce anomalies in time-series data. Techniques like co-simulation and multi-agent modeling and simulation have been explored in the literature for this purpose \cite{8340668}. However, these methods often face challenges in real-time systems, especially those with complex logic or multiple interconnected sensor elements. Researchers must critically assess the compatibility of these approaches with real-time systems and consider alternative methods for analyzing and defending against cyberattacks.

In addition to these approaches, several \textbf{time-series anomaly detection benchmarks} have been developed to aid researchers. For example, the \textbf{Secure Water Treatment (SWaT)} dataset \cite{goh2017dataset} includes real and simulated data from a water treatment plant, enabling studies on securing water treatment systems. Other popular benchmarks include the \textbf{SKAB} dataset \cite{skab} and the \textbf{Numenta Anomaly Benchmark (NAB)} \cite{AHMAD2017134}. While reproducing these benchmarks may be time-consuming, they offer an opportunity for researchers to evaluate their
anomaly detection algorithms and compare them against existing ones.

However, it is important to note that
these benchmarks may lack the robustness and scalability needed for certain research purposes; therefore, researchers should exercise caution when interpreting and generalizing their findings.

\section{\uppercase{Method}}
\label{mainSection1}

This chapter introduces SCART's architecture, beginning with a general description of the process for creating the layer used in most simulations. Subsequently, we present scenarios that guided our objectives during the design phase. These scenarios, based on the defined threat model, operate under the assumption that most modern computer systems are susceptible to cyber-attacks.

The primary objective of integrating the SCART layer into a simulator is to augment the functionality of an existing cycle-accurate or Digital Twin simulator, which closely mimics the target system's hardware and communication patterns. The proposed approach introduces the attack layer as an external modification to memory locations or communication channels, by integrating it into individual sub-units, their communication pathways, or the surrounding environment. The internal implementation of the SCART layer employs a discrete event simulation approach, modeling the system's behavior as a series of events that occur at distinct moments and signify changes in the system's state\cite{varga2001discrete}.

The layered architecture presented in this study provides a broad scope for investigating a wide range of threats, employing scenario definition encompassing both simple faults and complex, intricate attacks. Through parametric generation, the framework facilitates the production of numerous attacks across various threat categories. By partitioning the system into operational units and leveraging configuration files and parameters,  researchers can precisely define both nominal and abnormal behaviors. This inherent flexibility empowers researchers to conduct comprehensive simulations that faithfully depict the complex nature of potential threats.

\subsection{Architecture Overview}
\label{ArchitectureOverview}
As previously discussed in this chapter, the execution of the attack consists of a sequence of interconnected events and conditions. This approach is based on the SCART code layer, primarily on generating scenarios. It is crucial to highlight that these scenarios form the basis of our attack. Therefore, precisely defining each scenario that constitutes the intended attack is essential. This process requires a clear understanding of a scenario’s components and identifying its compatible parameters. By accomplishing this, we can proceed to the subsequent step, which involves  incorporating these scenarios into our layer, running them in the simulator, and effectively extracting the resulting information.
Within the SCART system, attacks are defined by three primary stages, all of which are pivotal for successful implementation:

\begin{enumerate}
    \item Defining the attack scenario: This stage involves formulating a detailed specification for the desired attack scenario. It entails determining the conditions under which the scenario should occur and outlining the corresponding actions to be executed. It necessitates a meticulous analysis of the target system and the desired outcome of the attack.

    \item Installing the scenario: Once the attack scenario is defined, it must be integrated into a suitable simulator environment. This process involves establishing the connections between the defined conditions, the start and end points, and the specified actions. In some cases, it also entails defining the interconnections between multiple scenarios. This integration enables the realistic emulation and execution of the attack scenario under controlled conditions.

    \item Running the scenario: After installing the scenario in the simulator, the subsequent step involves executing the attack scenario and observing its behavior closely. This process entails running the updated simulator when SCART is integrated and when the scenarios are installed. This stage facilitates the collection of valuable information regarding the impact, effectiveness, and potential countermeasures against the attack.
    
\end{enumerate}

    In the upcoming sections, we will delve into the specifics of each step in our implementation process.
    
\subsubsection{Define Scenario} \hfill
    
    To aid this exploration, we utilize three comprehensive lists to define and describe each scenario:
    
    \begin{itemize}
    \item  \( \mathcal{L}\text{isteners} = [sensor_1,sensor_2,\dots,sensor_n]\) \hfill \break
    
        This list represents the sensors we actively monitor. Our goal is to detect when a new message arrives for each sensor. We primarily use the PUSH methodology, but if unavailable, we resort to polling. Upon receiving an update for a given sensor $S_i$, we record its new value alongside previous values for historical tracking. To achieve this, we instantiate an object:
        \[ \mathcal{H}istory = \{s_1:[s_{11},\dots,s_{1t} ],s_2:[s_{21},\dots,s_{2t} ],\dots\} \]
  
         This object will record the system's state and memory starting at the execution's commencement.
         \\
         
        \item \(\mathcal{C}ondition=[c_1,c_2,\dots,c_p ]\) \hfill \break
        
        This list defines the conditions that must be met before an anomaly can be triggered. After every update in the History object, we verify if all specified conditions hold. Since we assume only one anomaly occurs per run, there is a time range $\Delta$t where all conditions hold together.
        Otherwise, the anomaly should not exist even in a real case. During this time, we will implement our exceptions. In other words, we will conclude the anomaly when any of the conditions are no longer satisfied. Furthermore, we can emphasize this aspect in the results whenever it occurs. \newline

        \item 
        \(\mathcal{A}ctions=[a_1,a_2,\dots,a_k ]\)\hfill \break

    This list specifies the actions or exceptions applied when all conditions are met. These actions introduce an anomaly by modifying system behavior. They may involve altering sensor values in the History object, modifying system memory, sending messages to sensors, or other operations. The length of this list is independent of the Listeners and Conditions lists, as exceptions may involve sensors unrelated to the monitored inputs and conditions.
            
        \end{itemize}

Having acquired a comprehensive understanding of our data structures, it is incumbent upon us to explain the operational intricacies of the architecture. In order to elucidate its functionality and versatility in simulating various attacks, we will present an example for that effectively define various scenarios. This instance illustrate the breadth of scenarios that can be defined within the architecture, thus demonstrating its ability to simulate diverse attacks and deviations. 
    
    \begin{itemize}
        \item GPS Spoofing Attack:  One exemplification entails the manipulation of the drone's position when it traverses a specific geographic location x0. Specifically, during this occurrence, the drone's position undergoes alteration by introducing a transient angular $\alpha$ deviation, which persists for milliseconds $t_0$. \\
        
        \( \mathcal{L}{isteners} = [sensor_1 = GPS]\)

        \(\mathcal{S}tart\mathcal{C}ondiction=[c_1 = \{GPS \approx x_0 \} ]\)
        
        \(\mathcal{E}nd\mathcal{C}ondiction=[c_2 = \{duration < t_0 \} ]\)

        \(\mathcal{A}ctions=[a_1 = \{duration = duration +1\} ,a_2 = \{GPS = \frac{duration}{t_0} \ast \alpha \ast GPS\} ]\)
        \end{itemize}

\subsubsection{Install Scenario} \hfill

Once the scenarios have been defined, the next step is their installation. Installing a scenario involves collecting all the lists and adapting them to a logical framework in which they can work together seamlessly. We utilize a Callback function that is triggered whenever a sensor updates its value to accomplish this. Depending on the specific requirements, different Callback functions may be used for individual sensors, or a shared function may be applied to a group of sensors. The Callback function is the interface that determines how the Condition and Action lists are utilized based on the information gathered from the Listeners stored in History.

The Callback function allows us to explore various implementation options and accommodate more complex attacks and scenarios. The following approach can be adopted as a vanilla example:
"Wait until all the starting conditions defined in the StartingCondition list are satisfied, and upon their completion, activate each anomaly specified in the Actions list. The execution continues until one of the EndingConditions is met, at which point the process halts."

This installation mechanism, coupled with the flexibility of the Callback function, enables the architecture to effectively orchestrate the execution of scenarios, allowing for intricate and sophisticated simulations of attacks and deviations.

\subsubsection{Running Scenario}
   The next phase in the implementation of an attack is running the scenario. In this phase, we adjust the architecture of our layer to the system’s simulator. Running the scenario will vary for each system, which requires customizing the SCART API to suit the specific simulator.

    Using the SCART API, we can configure how the system’s sensors are read and create a list of listeners for our layer. Furthermore, we can control how the layer updates sensor values after the scenario is activated.
    
    This phase plays a crucial role in mitigating supply chain attacks and ensuring persistence after the initial foothold. For a more comprehensive exploration of embedded system threats, including supply chain attacks and persistence techniques.

\subsection{Attack Scenarios}
         \label{sec:AttackSenario}
        
The threat model used to design SCART presupposes that the attacker may be external to the system or use a Hardware Trojan Horse (HTH), meaning that the attacker has already successfully conducted a classic stealth attack. These assumptions allow us to simulate a broader attack surface encompassing diverse attack vectors.

To support this, SCART implements the attack layer so that the attacker waits for a specific trigger before launching the attack. This implementation allows us to support a wide range of attacks, including:
            
\begin{itemize}
    \item \textbf{Attacking the external control system}
    
    Here, we present an overview of the threats that arise from the system's external interfaces, referred to as controller or management interfaces. Examples of such interfaces include remote control, command-and-control positions, and management positions. These threats can be executed by injecting malicious code into the controller’s firmware or software and, in some cases, by developing alternative communication interfaces for the control system.
    
    \item  \textbf{System firmware attacks}
   
  Firmware serves as the core of real-time systems, making it a primary target for attacks. The attacker receives the telemetry values directly from the sensors. Our code package will be included as a component of the real-time system itself to provide an attack mechanism in such a scenario. This attack vector may be accomplished by adding the necessary code to the real-time system firmware at the point when the system sensors synchronize.
    
    \item  \textbf{Attacks through the internal communication interfaces}
    
    As we presented at the beginning of the chapter, we divide our real-time system into sub-modules that communicate with each other. This communication is a widespread threat to systems, as mentioned in many publications \cite{thing2016autonomous},\cite{papp2015embedded}, \cite{parkinson2017cyber}. Hence, the integration of an attack mechanism in such a situation will be integrating our layer in the communication interfaces between the sub-modules. This integration can be achieved by adding the code to the internal communication components as a proxy or indirectly injecting relays.

    \item  \textbf{Attack through the environment}
    
   A huge part of developing a simulation is adapting it to the real world. As we all know, real-time systems receive input from their surroundings in a non-communicative manner. Accordingly, in most simulators, it is necessary to simulate the elements of the environment, such as wind, obstacles, sun, and more. Many studies presented the threat using environmental elements, i.e., Patch Attacks and Adversarial attacks on visual odometry systems \cite{nemcovsky2022physical}. Therefore, we will integrate our layer as part of the simulator and thus use the simulator and inject the desired attacks into the simulation environment.
    \end{itemize}
Although SCART's capabilities extend to all the scenarios described, the experiments and evaluations in this work primarily focus on attacks through communication interfaces.

\section{\uppercase{Simulation Methodology}}
\label{mainSection2}
This chapter outlines the process of integrating SCART into simulators, emphasizing its versatility and applicability. By seamlessly incorporating SCART into an existing simulation framework, researchers can analyze and execute a wide range of attack scenarios in a controlled and realistic environment. Additionally, this chapter explores the application of SCART to specific attack scenarios, building upon the foundation established in previous chapters. Through this integration, researchers can systematically evaluate the impact of various attack scenarios within their chosen simulation environment.

To seamlessly integrate SCART into an existing simulator, it is crucial to identify the core pathway that connects the input to the output within the chosen simulation framework. Specifically, this involves locating the Main function responsible for linking sensor data to system behavior or navigation commands. To facilitate this connection, a proxy will be developed to serve as an intermediary between the input and output components.
Once the proxy is established, SCART will be integrated with it. SCART will process the inputs received from the proxy, modify them according to the installed scenarios, and generate updated inputs based on these modifications. Consequently, the output from SCART will act as the attacked input for the simulator, enabling controlled testing of various attack scenarios.

\begin{figure}[!t]
    \centering
    \begin{subfigure}[b]{0.49\linewidth}
        \includegraphics[width=\linewidth,height=0.1\textheight]{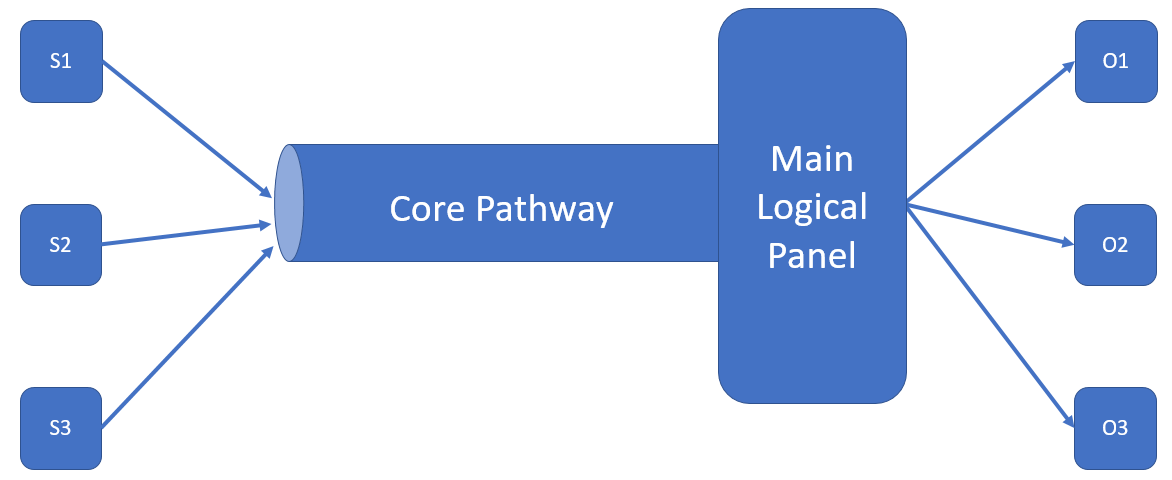}
        \caption{Before integration.}
    \end{subfigure}
    \hfill
    \begin{subfigure}[b]{0.49\linewidth}
        \includegraphics[width=\linewidth,height=0.1\textheight]{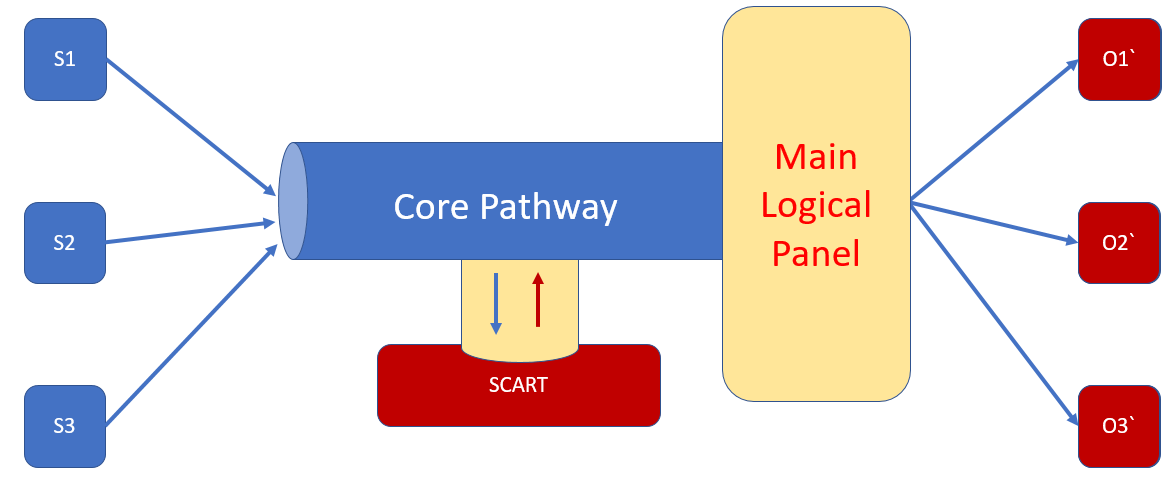}
        \caption{After integration.}
    \end{subfigure}
    \caption{How to integrate SCART into an existing simulator. The left image shows the system diagram before integration, and the right image shows it after integration.}
    \label{fig:PipeAndMore}
\end{figure}

A helpful analogy to clarify this process is to compare it to a home electrical system. Instead of individually altering each power consumer, attention is directed to a central junction box or main power panel. A "man in the middle" is then introduced between the panel and the consumers' boxes, effectively influencing the activities of the consumers. A visual representation of this process can be found in Fig.\ref{fig:PipeAndMore}.

To validate the proposed solution, a drone system based on PX4 firmware was selected for experimentation. Drones provide an excellent example of real-time systems due to their precise, parameter-based control systems. Moreover, the drone system operates on a real-time cycle clock, where failure to meet time constraints could compromise the system's overall performance. In simpler terms, the process of adding the SCART logic to an existing simulation system involves the following steps:

\begin{enumerate}
    \item \textbf{Identify the Core Pathway:} Locate the central Main function within the chosen simulator that establishes the connection between the sensor input and system behavior or navigation commands.
    
    \item \textbf{Develop a Proxy:} Create an intermediary proxy to bridge the input and output components of the simulator, enabling the establishment of a connection with SCART.
    
    \item \textbf{Link SCART to the Proxy:} Establish a linkage between SCART and the proxy, allowing SCART to receive inputs from the proxy and modify them according to the predefined attack scenarios.
    
    \item \textbf{Generate "Updated" Inputs:} Within SCART, generate "updated" inputs based on the specified attack scenarios, transforming the modified inputs from the proxy into attack-oriented inputs for the simulator.
\end{enumerate}

 \subsection{Attack via communication Interfaces}
\label{AttackCommunicationInterfaces}
Most PX4-based systems consist of a flight controller and a mission computer, which are interconnected. The flight controller manages the aircraft's physical behavior, utilizing sensors and algorithms to control its movements. The mission computer serves as the primary interface for the user, enabling them to program and direct the drone's tasks. Typically, the mission computer is a separate module. These two modules communicate through the ``MAVLink'' protocol. This structure allows us to demonstrate an attacker positioned within the computer's communication interface, thus performing a Man-in-the-Middle attack.

To simulate the drone, we chose to use the Gazebo simulator. Gazebo \cite{kaur2021survey} is an open-source 3D robotics simulator that supports the simulation of sensors and operator control. It also allows the creation of different variables and specific environmental conditions. Additionally, Gazebo enables modifications to the drone's environment and the addition of custom objects. To supplement Gazebo, we leveraged the MAVROS package \cite{lee2021robot} to establish communication between the PX4 autopilot and the mission computer. MAVROS served as the MAVLink node for the Robot Operating System (ROS), providing access to critical flight and state information from the drone. Using this setup, we demonstrate a simple attack scenario through the internal communication interfaces.

After selecting the most relevant simulator, we divided the main logic into separate running modules. Specifically, we integrated the PX4 hardware with the simulator and placed them in a distinct module, which we named "Simulator." We then created a new "mission computer" module that takes the flight controller's sensor sequence and passes it to the mission computer. The flight controller receives the desired mission as a parameter at the task's start and calculates the drone's movement based on the sensor sequence. The mission computer is named "Navigator." Consequently, we now have two distinct modules, "Simulator" and "Navigator," that communicate using the ``MAVLink'' protocol.

Next, we added our layer's logic, our implementation involved building a python-based demo\footnote{The link has been omitted to preserve the anonymity of the submission} that intercepts sensor data before it enters the mission computer's logic. This demo executes predefined attack scenarios and feeds the processed sensor data to the mission computer. Integration with MAVROS enabled us to incorporate this functionality with minimal additional effort. We remapped the sensors so that our layer receives the sensor values before they reach the mission computer. In this way, our layer acts as an attacker within the communication module, performing a Man-in-the-Middle attack \cite{cao2020survey}.

\begin{figure}[htbp]
  \centering
  \begin{subfigure}[b]{0.49\linewidth}
    \includegraphics[width=\linewidth]{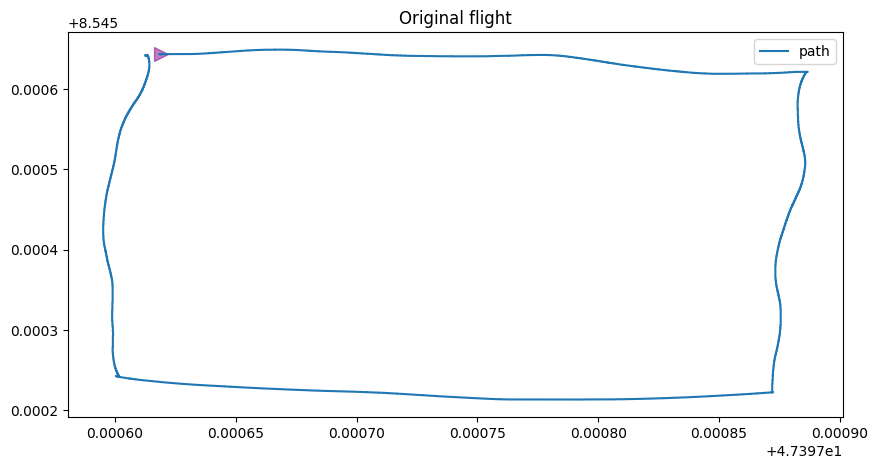}
    \caption{Before the attack.}
  \end{subfigure}
  \hfill
  \begin{subfigure}[b]{0.49\linewidth}
    \includegraphics[width=\linewidth]{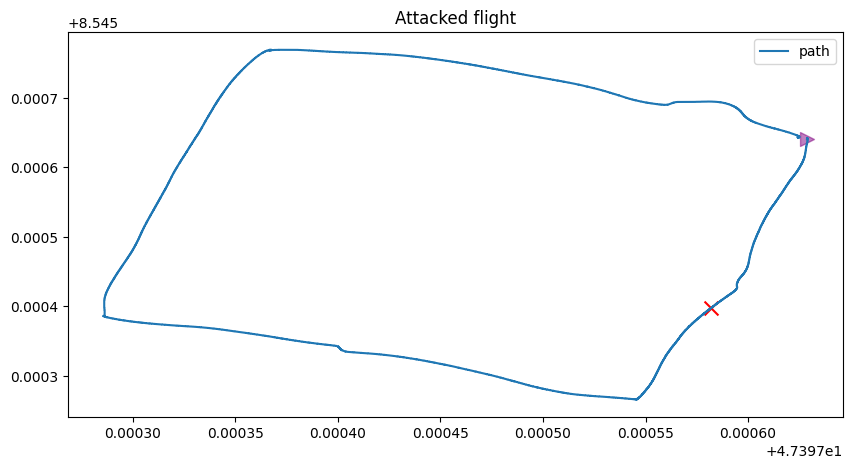}
    \caption{After the attack.}
  \end{subfigure}
  \caption{The flight path before and after the attack.}
  \label{fig:AddWhiteNoisePath}
\end{figure}

In this phase, the installation of the scenarios and their execution logic was implemented following the remapping of the sensors. First, the logic for installing the vanilla scenario was developed, as previously described. Subsequently, we created the corresponding list of listeners based on the remapped sensors. In our setup, the listeners receive sensor values through a ROS object \cite{quigley2009ros} called "Subscriber," to which the sensor name and callback function are passed and operated using the PUSH method. The listeners wait until the reception of a new sensor record triggers the callback function. In this example, the subscriber's values are stored in the history structure, as detailed in Section \ref{ArchitectureOverview}. Once our layer has implemented the installed scenarios, the new sensor values are sent to the mission computer using the "Publisher" object. At this stage, the mission computer receives an update from the flight controller, executes the navigation algorithm, and sends an anomaly notice to the flight controller. In Fig. \ref{fig:AddWhiteNoisePath}, we present the flight path before and after the attack as part of the experimental results.

\begin{figure}[ht]
\includegraphics[width=\columnwidth]{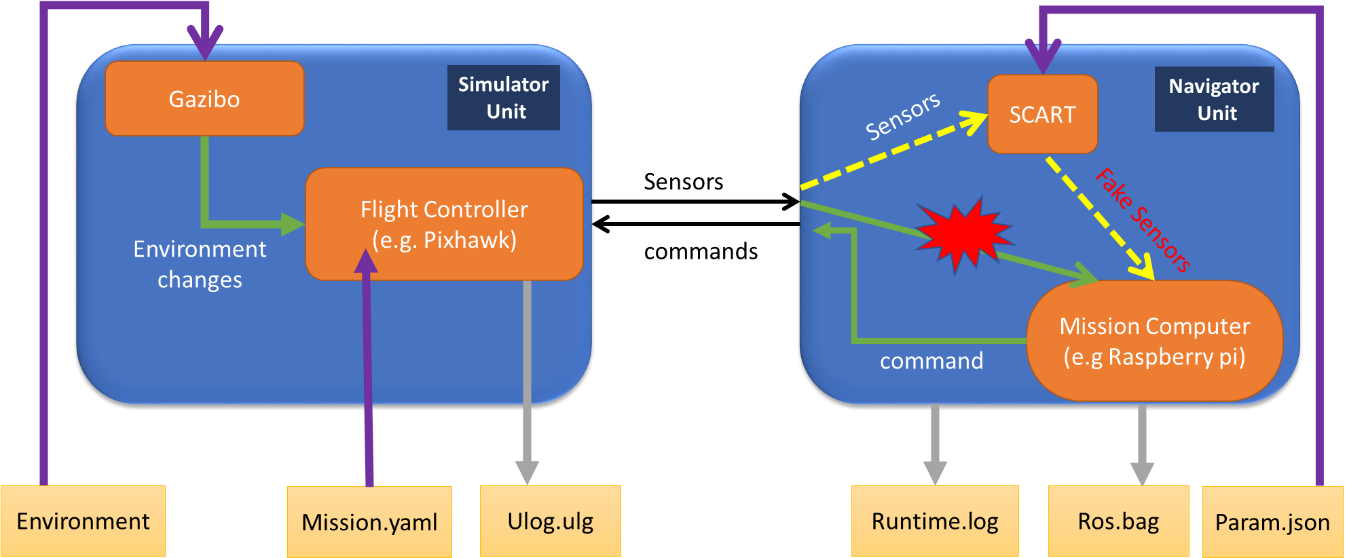}
\centering
\caption{Environment layout for internal communication interfaces scenario}
\label{fig:archOverview}
\end{figure}

To facilitate a more comprehensive understanding of our experimental setup, we present in Fig. \ref{fig:archOverview} a diagrammatic representation of the communication channels involved. Specifically, the green arrows represent the original communication paths, while the yellow arrows indicate the changes made by our setup. The gray arrows show the setup's output, while the purple arrows signify its input. Finally, the black arrows represent the communication interface between the simulator and the Navigator.

    \subsection{Dataset and Collection}
    \label{Dataset}

This section explains the dataset creation methodology using our SCART system, highlighting its ability to efficiently and robustly create datasets and benchmarks. Our method allows exporting all sensor records from each system iteration. With the help of our layer, we can export the history object at the end of each run, generating a comprehensive database, especially when documentation interfaces for sensor values are not available.

We demonstrate how to export a dataset from the simulators, where the data is a product of the simulators, and our added functionality is the creation of cyber attacks or malfunctions proactively and parametrically. Most of the logs are flight recordings from the Navigator and Simulator, represented by yellow squares in Fig.\ref{fig:archOverview}.

To ensure the relevance of the exported data, we first verified our simulator’s accuracy by running it in three Gazebo modes: hardware-in-the-loop (HITL), software-in-the-loop (SITL), and accurate flight mode. We compared the logs obtained from each mode and found that the Gazebo simulation accurately represents flight and produces valid environmental values for use.

Dataset creation is essential for both training and testing, particularly when many machine learning-based anomaly detection methods are unsupervised, learning solely from non-anomalous data and tested on anomalous data, as demonstrated in Section \ref{mainSection3}. In this regard, our system provides the first high-quality testing solution for a wide range of real-time anomaly detection algorithms.

\begin{figure*}[htbp]
  \centering
  \begin{subfigure}[b]{0.32\textwidth}
    \includegraphics[width=\linewidth]{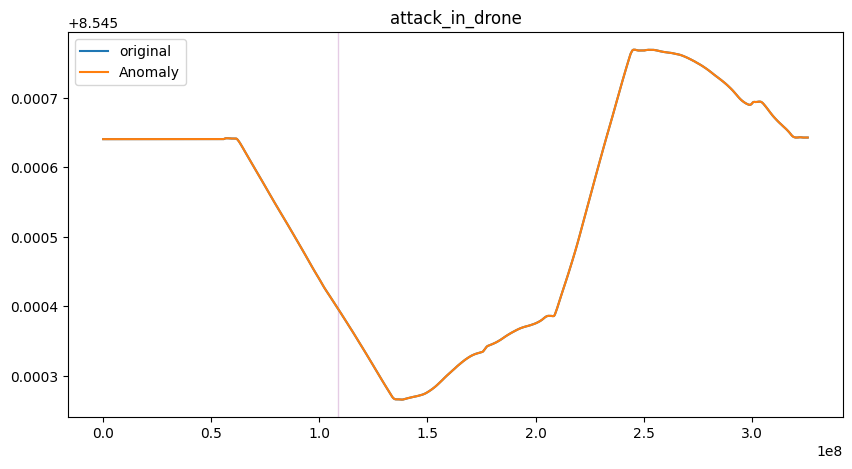}
  \end{subfigure}
  \hfill
  \begin{subfigure}[b]{0.32\textwidth}
    \includegraphics[width=\linewidth]{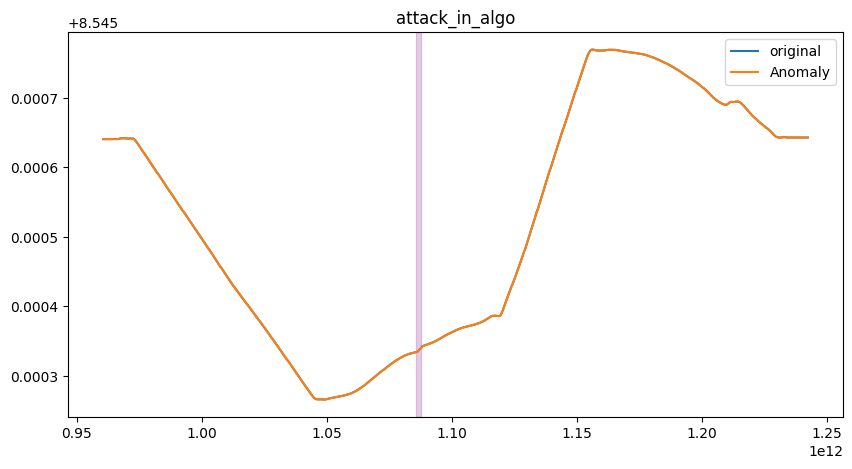}
  \end{subfigure}
  \hfill
  \begin{subfigure}[b]{0.32\textwidth}
    \includegraphics[width=\linewidth]{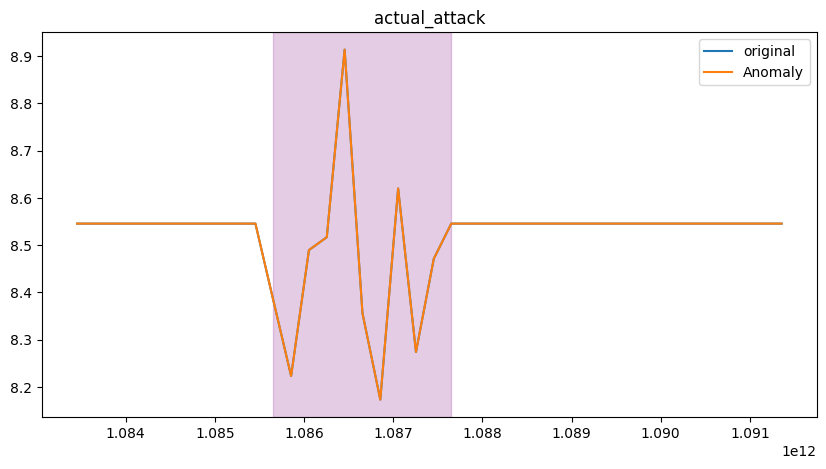}
  \end{subfigure}
  \caption{Example of AddWhiteNoise attack across three scenarios. The
left image shows the actual appearance of the AddWhiteNoise attack, while the middle and right images depict the sensor readings in the Navigator and
Simulator, respectively.}
  \label{fig:LatAddWhiteNoise}
\end{figure*}

\subsubsection{Dataset Structure}\hfill    

The dataset has been carefully organized into separate folders to ensure easy access and efficient utilization. The first folder contains all files collected during the scenario, including logs and configuration files for the mission. The second folder holds sensor files generated through automated tools and various manipulations of the raw data. Additionally, a third folder contains two CSV files: one that combines all the sensor values into a single table, and another that includes the most frequently sampled sensor values. Both CSV files also feature a field to flag anomalies in each entry. This methodical organization facilitates efficient analysis and enables the identification of anomalies. The complete list of sensors can be found in the following CSV file:

\begin{table}[!ht]
    \centering
    \resizebox{\linewidth}{!}{ % Automatically scales the table to fit the column width
    \begin{tabular}{|c|c|c|c|c|c|c|}
    \hline
    \textbf{Index} & \textbf{Timestamp} & \textbf{Is\_Anomaly} & \textbf{...} & \textbf{Latitude} & \textbf{Longitude} & \textbf{...} \\ \hline
    0    & 0         & 0   & ... & 0         & 0        & ... \\ \hline
    ...  & 0         & 0   & ... & 0         & 0        & ... \\ \hline
    8963 & 108604000 & 0   & ... & 47.397404 & 8.545335 & ... \\ \hline
    8964 & 108616000 & 1   & ... & 47.397403 & 8.545336 & ... \\ \hline
    ...  & ...       & ... & ... & ...       & ...      & ... \\ \hline
    \end{tabular}
    }
    \caption{Example of our dataset structure}
\end{table}

\begin{itemize}
    \item The first column is a running index.
    
    \item The second column, called "timestamp," represents the time from the start of the run in microseconds.
    
    \item The third column, called "is-anomaly," is a Boolean field indicating whether there is an anomaly at this timestamp.
    
    \item From the fourth column onward are the values of the measured sensors.
\end{itemize}

\section{\uppercase{Evaluation}}
\label{mainSection3}
To safeguard real-time systems against vulnerabilities and cyber-attacks, it is crucial to differentiate between acceptable system behavior and anomalous behavior. Previous methods for anomaly detection have been extensively surveyed by \cite{chandola2009anomaly}. However, one of the main challenges faced by current anomaly detection techniques is defining normal behavior, given the complexity and variability of real-time systems, along with the lack of well-defined metrics for normality. Despite this, real-time systems provide additional data that can help address this challenge. For example, real-time systems are typically cycle-based, where deterministic execution times are essential to ensure each computing unit completes its task within specified time constraints \cite{ben2006principles}.

During the experimental phase, a controlled environment was set up to faithfully replicate the operational setup of the drone, facilitating a wide range of experiments both indoors and outdoors. These experiments considered various factors, such as weather changes, geographical variations, and visibility conditions. Environmental models, including physical objects, were also incorporated, all aimed at simulating realistic flight conditions.

Simultaneously, the laboratory’s simulation environment, including Gazebo, was aligned with the experimental setup to ensure seamless integration between simulations and real-world scenarios. This alignment provided a comprehensive evaluation of the SCART solution. These experimental flights, along with software and hardware-in-the-loop tests, were pivotal in assessing SCART’s robustness and effectiveness under diverse real-world conditions.

Integrating real-time data from drones into the simulation system provided valuable insights into SCART’s applicability and reliability, demonstrating its potential to enhance security and improve the performance of real-time systems.

In the next stage of our research, we evaluate the SCART method by examining the architecture developed, incorporating SCART as described in Section \ref{Dataset}. The model's effectiveness is assessed by comparing sensor readings against a control run with no exceptions, primarily focusing on common attacks such as GPS spoofing \cite{tippenhauer2011requirements}.

In the remainder of this chapter, we employ different anomaly detection algorithms. Through the evaluations performed, we demonstrate that SCART can enhance the performance of detection algorithms, especially when generating anomalous data or attacking critical systems proves challenging.

\begin{figure*}[htbp]
  \centering
  \begin{subfigure}[b]{0.32\textwidth}
    \includegraphics[width=\linewidth]{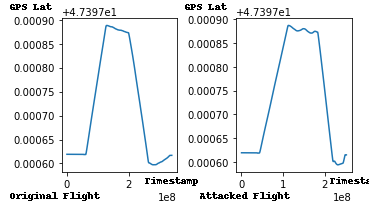}
  \end{subfigure}
  \hfill
  \begin{subfigure}[b]{0.32\textwidth}
    \includegraphics[width=\linewidth]{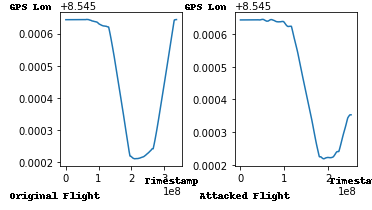}
  \end{subfigure}
  \hfill
  \begin{subfigure}[b]{0.32\textwidth}
    \includegraphics[width=\linewidth]{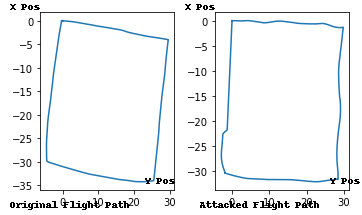}
  \end{subfigure}
  \caption{Changes in the behavior of position-related sensors caused by GPS spoofing with random modifications, generated by SCART.}
  \label{fig:random_change}
\end{figure*}

\begin{figure*}[htbp]
  \centering
  \begin{subfigure}[b]{0.32\textwidth}
    \includegraphics[width=\linewidth]{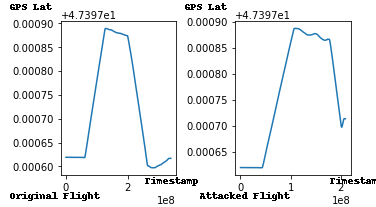}
    % \caption{\footnotesize{ Latitude sensor values during regular flight (left) and during the attacked flight (right).}}
  \end{subfigure}
  \hfill
  \begin{subfigure}[b]{0.32\textwidth}
    \includegraphics[width=\linewidth]{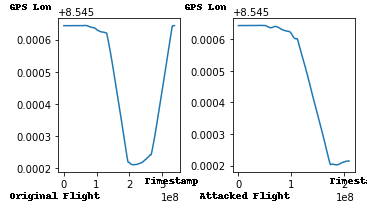}
    % \caption{\footnotesize Longitude sensor values during regular flight (left) and during the attacked flight (right).}
  \end{subfigure}
  \hfill
  \begin{subfigure}[b]{0.32\textwidth}
    \includegraphics[width=\linewidth]{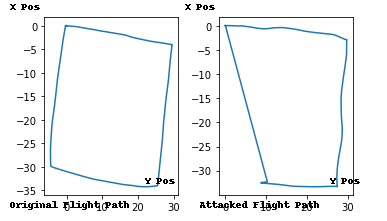}
    % \caption{Drone path (x, y coordinates) during regular flight (left) and attacked flight (right).}
  \end{subfigure}
  \caption{Changes in the behavior of position-related sensors caused by GPS spoofing with duplicate attack patterns, generated by SCART}
  \label{fig:douplicate}
\end{figure*}

\label{mainSection4}
\subsection{Experiments}

We conducted comprehensive experiments on the system within an attack scenario, as detailed in Section \ref{AttackCommunicationInterfaces}. These experiments were carried out in a simulation environment designed to meet the project's specific requirements, comprising two operational units. The first unit, referred to as the "Simulator," was developed by integrating the Gazebo simulator with custom modifications and adjustments, specifically designed to meet the project’s needs. This unit included various internally developed modules to enable flight simulations in a virtual environment while running the drone’s firmware. The second unit, referred to as the "Navigator," emulated the algorithmic aspect of the system by processing sensor data from the Simulator and executing the necessary computations to generate corresponding commands. Both units, the Simulator and Navigator, functioned in tandem, remaining synchronized during both simulated and real-world flight scenarios.

The primary aim of our experiments was to evaluate the performance of existing algorithms in identifying specific anomalies. To achieve this, we created a simulator unit tasked with executing a predetermined mission, spanning from takeoff to landing, and involving a series of sequential tasks. For comparison, we performed 100 simulations of the system with no attacks, following a predefined 30x30 square flight path. 

The primary aim of our experiments was to evaluate the performance of existing algorithms in identifying specific anomalies. To achieve this, we created a simulator unit tasked with executing a predetermined mission, spanning from takeoff to landing and involving a series of sequential tasks. For comparison, we performed 100 simulations of the system with no attacks, following a predefined 30x30 square flight path.

Throughout the experiments, we applied various sensor manipulation techniques, resulting in a total of 2048 potential attack scenarios. Subsequently, we evaluated the effectiveness of various anomaly detection methods in identifying these attacks and recorded the results in tables. Our data is structured in columns, with each column representing time-series data from a specific sensor. The criteria used to determine whether an anomaly detection method successfully identified an attack will be elaborated upon in subsequent sections.

As depicted in Figure \ref{fig:LatAddWhiteNoise}, we present the outcomes of the experiments for each attack type. The left image shows the actual appearance of the AddWhiteNoise attack, while the middle and right images depict the sensor readings in the Navigator and Simulator, respectively. Notably, the red markings indicate the time at which the attack was initiated, providing a clear visual representation of its impact.

Figures \ref{fig:random_change} and \ref{fig:douplicate} illustrate examples of changes in the behavior of sensors related to position caused by GPS spoofing generated by SCART, demonstrating the effects of random changes and duplicate attacks, respectively.

   \subsection{Anomaly Detection Results}
In this section, we assess the performance of our API in detecting anomalies in time series data. Anomalies refer to patterns that deviate from normal behavior, and detecting such deviations is crucial in many domains, particularly in cybersecurity. We generated three types of datasets for evaluation: normal flights, flights attacked in simulation mode, and flights attacked in CSV mode. The performance of the API was evaluated by calculating true positives (TP) and true negatives (TN). A TP represents a case where the algorithm correctly detects an anomaly in an attacked flight. A TN represents a case where the algorithm correctly identifies that no anomaly exists in a normal flight.

We compared our approach against several existing anomaly detection techniques. First, we evaluated the ``Telemanom'' algorithm, presented by \cite{hundman2018detecting}, which uses Single-Channel Models of LSTM followed by unsupervised thresholding to detect anomalies in telemetry data streams. Additionally, we considered the ``Tsbitmaps'' algorithm, based on \cite{Wei2005AssumptionFreeAD} and \cite{Kumar2005TimeseriesBA}, which identifies outliers by comparing the frequency of local regions in adjacent subsequences using bitmap representation and squared Euclidean distance, with incremental updates at each time step.

We also explored the effectiveness of simple statistical methods, as discussed in \cite{wu2021current}, including averaging, standard deviation, variance, and moving averages, to detect anomalies. For our evaluation, we defined an anomaly point as one that exceeds multiple statistical thresholds simultaneously.

Finally, we examined deep learning models for anomaly detection, specifically ``deepod'' \cite{deepod}, as presented in \cite{pang2019deep}. We trained six models using our normal flight data and evaluated their ability to detect anomalous flights. Each model was trained to learn the patterns of normal flights, enabling it to distinguish between normal and anomalous flight data.

    \begin{table}[!ht]
        \begin{center}
\begin{tabular}{|l|l|l|l|}
\hline
  & \multicolumn{2}{l|}{TP} & TN        \\ \hline
  & Simulation     & CSV    & No Attack \\ \hline
    Telemanon   &  58.2\% & 89.6\% & 90\%\\\hline
    Tibstib & 42\% & 86.2\% &  100\%  \\ \hline
    Statistical   & 59.7\% & 86.2\% & 100\% \\ \hline
    DeepSVDD   & 100\% & 100\% & 90\% \\\hline
    RDP   & 100\% & 100\% &  100\% \\  \hline
    RCA   & 100\% &  100\%  &100\%\\ \hline
    GOAD   & 100\% &  100\%  &100\%\\ \hline
    NeuTraL & 100\% & 100\% & 100\%  \\ \hline
    \end{tabular}
    \end{center}
    \caption{Anomaly detection results}
    \end{table}

The following table provides an overview of the performance of various anomaly detection algorithms on data generated using SCART. As shown in the table, most of the algorithms successfully identified anomalies in the data.

\begin{figure}[htbp]
  \centering
  \begin{subfigure}[b]{0.49\linewidth}
    \includegraphics[width=\linewidth]{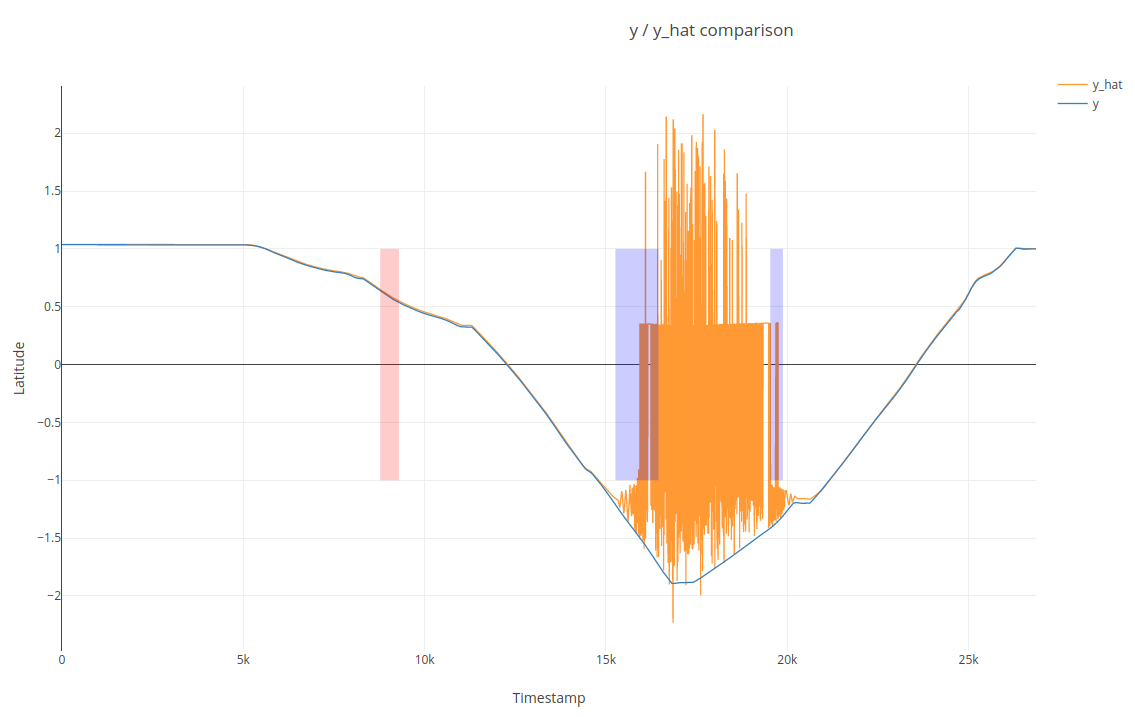}
    \caption{Latitude detection.}
  \end{subfigure}
  \hfill
  \begin{subfigure}[b]{0.49\linewidth}
    \includegraphics[width=\linewidth]{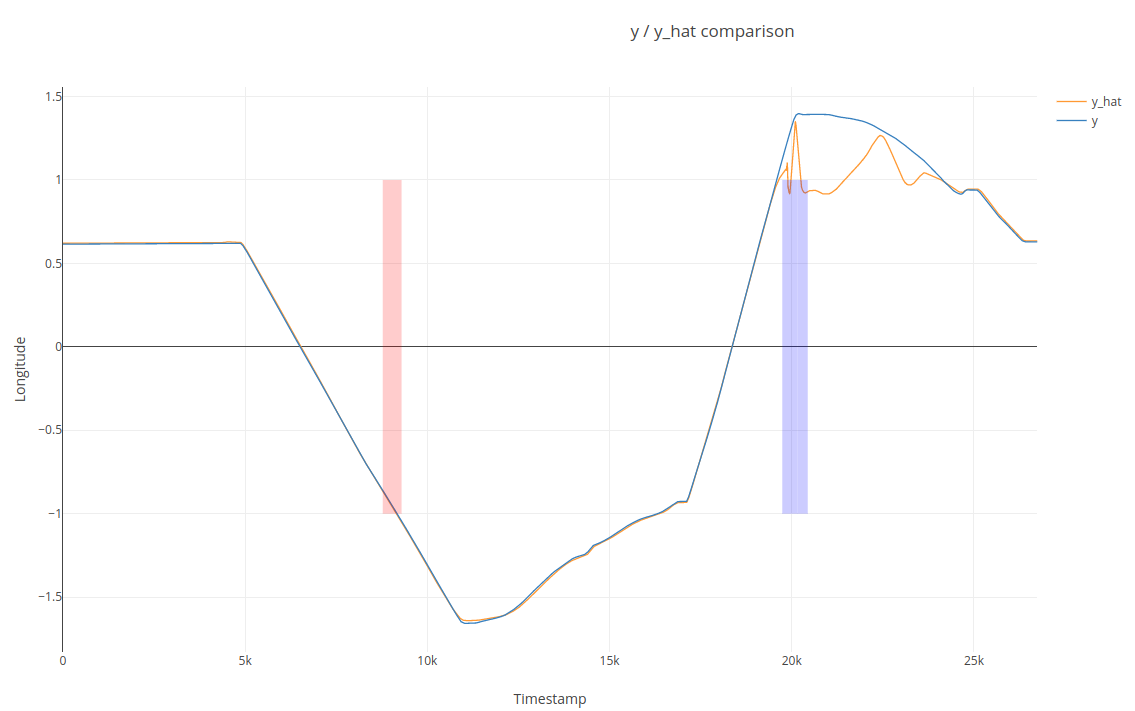}
    \caption{Longitude detection.}
  \end{subfigure}
  \caption{Result of trying to detect the attack with Telemanom.}
  \label{fig:TelemanomDetection}
\end{figure}

Additionally, Figure \ref{fig:TelemanomDetection} presents the results of running the Telemanom algorithm on one of our experiments. The left and right images display the latitude and longitude sensor readings, respectively. Each image includes two graphs: the blue graph represents the output generated by SCART with the AddWhiteNoise attack activated, while the orange graph shows the predictions made by the Telemanom algorithm for the anomalous flight. The purple rectangles highlight the detected anomaly ranges, indicating the duration of the attack, while the red rectangles mark the period of the attack generated by SCART. Overall, our results demonstrate that the Telemanom algorithm successfully detected the anomaly introduced by SCART, showcasing the effectiveness of SCART in simulating cyber attacks and supporting the development of robust anomaly detection techniques.

\section{Conclusions and Discussion}  
\label{discussion}

This paper demonstrates the utility of the SCART framework for enhancing simulation environments in real-time systems, particularly in addressing reliability-related events and advanced cybersecurity attacks. By integrating the proposed SCART layer into existing simulation platforms, it is possible to replicate both random component or subsystem failures and sophisticated security threats. SCART’s effectiveness was validated by training machine learning algorithms to govern a drone’s flight control system, achieving higher accuracy and significantly reduced false-positive rates.

Our experiments further confirmed SCART’s capability to generate realistic cyber-attacks, with multiple anomaly detection methods successfully distinguishing between normal and compromised flights. These results highlight SCART’s ability to simulate diverse attack scenarios, providing a robust foundation for ongoing research in anomaly detection.

The API and preliminary findings presented in the evaluation chapter establish a basis for further exploration of SCART-based solutions. Our research shows that SCART can generate high-quality datasets of cyber-attacks, essential for training deep learning models aimed at real-time anomaly detection with minimal false positives. Additionally, SCART provides a reliable means to evaluate and compare different anomaly detection algorithms, making it a powerful resource for algorithm refinement.

In summary, the SCART framework represents a significant advancement in real-time systems and cybersecurity. It offers a versatile and cost-effective tool for simulating complex systems and sophisticated cyber-attacks. We hope this work inspires further exploration of SCART’s capabilities and applications, contributing to more effective and reliable cybersecurity solutions.

\bibliographystyle{apalike}
{\small
\bibliography{example}}

\end{document}